\documentclass[journal=nalefd,manuscript=letter]{achemso}

\usepackage[version=3]{mhchem} % Formula subscripts using \ce{}
\usepackage[T1]{fontenc}       % Use modern font encodings
\usepackage{color}
\setkeys{acs}{articletitle=true}
\setkeys{acs}{maxauthors = 0}
\def\be{\begin{eqnarray}}   
\def\ee{\end{eqnarray}}
\def\vecb{\mathbf}   
\email{kanelson@mit.edu}

\author{Jiaojian Shi$^\bigtriangleup$}
\affiliation[MITChem]{Department of Chemistry, Massachusetts Institute of Technology, Cambridge, MA 02139, USA}

\author{Edoardo Baldini$^\bigtriangleup$}
\affiliation[MITPhys]{Department of Physics, Massachusetts Institute of Technology, Cambridge, MA 02139, USA}

\author{Simone Latini}
\affiliation[MPI]{Max Planck Institute for the Structure and Dynamics of Matter, Center for Free Electron Laser Science, 22761 Hamburg, Germany}

\author{Shunsuke A. Sato}
\affiliation[Tsukuba]{Center for Computational Sciences, University of Tsukuba, Tsukuba 305-8577, Japan}
\alsoaffiliation[MPI]{Max Planck Institute for the Structure and Dynamics of Matter, Center for Free Electron Laser Science, 22761 Hamburg, Germany}

\author{Yaqing Zhang}
\affiliation[MITChem]{Department of Chemistry, Massachusetts Institute of Technology, Cambridge, MA 02139, USA}

\author{Brandt C. Pein}
\affiliation[MITChem]{Department of Chemistry, Massachusetts Institute of Technology, Cambridge, MA 02139, USA}

\author{Pin-Chun Shen}
\affiliation[MITEE]{Department of Electrical Engineering and Computer Science, Massachusetts Institute of Technology, Cambridge, MA 02139, USA}

\author{Jing Kong}
\affiliation[MITEE]{Department of Electrical Engineering and Computer Science, Massachusetts Institute of Technology, Cambridge, MA 02139, USA}

\author{Angel Rubio}
\affiliation[MPI]{Max Planck Institute for the Structure and Dynamics of Matter, Center for Free Electron Laser Science, 22761 Hamburg, Germany}
\alsoaffiliation[San]{Nano-Bio Spectroscopy Group, Departamento de Fisica de Materiales, Universidad del Pa\'is Vasco, 20018 San Sebast\'ian, Spain}
\alsoaffiliation[NYC]{Center for Computational Quantum Physics, Simons Foundation Flatiron Institute, New York, NY 10010 USA}

\author{Nuh Gedik}
\affiliation[MITPhys]{Department of Physics, Massachusetts Institute of Technology, Cambridge, MA 02139, USA}

\author{Keith A. Nelson}
\affiliation[MITChem]{Department of Chemistry, Massachusetts Institute of Technology, Cambridge, MA 02139, USA}
\email{kanelson@mit.edu}

\title[]
  {Room Temperature Terahertz Electroabsorption Modulation by Excitons in Monolayer Transition Metal Dichalcogenides}

\abbreviations{}

\begin{document}

%\begin{tocentry}
%
%Some journals require a graphical entry for the Table of Contents.
%This should be laid out ``print ready'' so that the sizing of the
%Inside the \texttt{tocentry} environment, the font used is Helvetica
%8\,pt, as required by \emph{Journal of the American Chemical
%The surrounding frame is 9\,cm by 3.5\,cm, which is the maximum
%permitted for  \emph{Journal of the American Chemical Society}
%graphical table of content entries. The box will not resize if the
%content is too big: instead it will overflow the edge of the box.
%
%This box and the associated title will always be printed on a
%separate page at the end of the document.
%
%\end{tocentry}

\newpage
\begin{abstract}
The interaction between off-resonant laser pulses and excitons in monolayer transition metal dichalcogenides is attracting increasing interest as a route for the valley-selective coherent control of the exciton properties. Here, we extend the classification of the known off-resonant phenomena by unveiling the impact of a strong THz field on the excitonic resonances of monolayer MoS$_2$. We observe that the THz pump pulse causes a selective modification of the coherence lifetime of the excitons, while keeping their oscillator strength and peak energy unchanged. We rationalize these results theoretically by invoking a hitherto unobserved manifestation of the Franz-Keldysh effect on an exciton resonance. As the modulation depth of the optical absorption reaches values as large as 0.05 dB/nm at room temperature, our findings open the way to the use of semiconducting transition metal dichalcogenides as compact and efficient platforms for high-speed electroabsorption devices.
\end{abstract}
\newpage

Monolayer semiconducting transition metal dichalcogenides (TMDs) have revolutionized the field of optoelectronics during the last decade \cite{schaibley2016valleytronics}. Their almost ideal two-dimensional nature and their weak dielectric screening allow these materials to host strongly bound excitons with a unique character among all direct bandgap semiconductors. These excitons can be described within the Wannier-Mott regime, yet the large Coulomb interaction limits the extension of the electron-hole correlation only to several lattice periods \cite{qiu2013optical}. In addition, excitonic optical absorption in TMDs obeys chiral selection rules that depend on the valley of the Brillouin zone in which the electronic states building up the excitons reside. It is the coexistence of large oscillator strengths \cite{zhang2014absorption}, relatively long coherence lifetimes \cite{hao2016direct, selig2016excitonic}, and valley-selective control \cite{kim2014ultrafast, sie2015valley, sie2017large, sun2017optical} that has made excitons in TMDs an attractive platform for opto-valleytronic applications at room temperature. This, in turn, has stimulated increasing efforts to manipulate the exciton properties on an ultrafast timescale by means of optical excitation.

On one side, a vast literature has clarified the subtle interplay of single-particle and many-body effects (e.g., phase-space filling, Coulomb screening, bandgap renormalization, exciton-exciton interaction...) that emerge upon above-gap photoexcitation and persist until photocarrier recombination is complete \cite{mai2013many, sun2014observation, sie2015intervalley, schmidt2016ultrafast, pogna2016photo, mahmood2017observation, ruppert2017role}. On the other hand, only a few works have explored the off-resonant (i.e. below-gap) excitation regime. In this limit, intriguing nonlinear phenomena stem from the coherent dressing that the exciton experiences in the presence of a slowly-varying periodic photon field \cite{de2016monitoring}, such as that provided by short laser pulses in the mid-to-far infrared spectral range. Notable examples comprise the ac (optical) Stark effect \cite{kim2014ultrafast, sie2015valley, sim2016selectively, lamountain2018valley, yong2018biexcitonic}, the Bloch-Siegert shift \cite{sie2017large}, and the Autler-Townes splitting \cite{yong2019valley}. Yet, still unexplored are the nonlinear optical effects that excitons in TMDs show in response to an intense perturbation at terahertz (THz) frequencies. Unlike in semiconductors with weakly bound excitons \cite{danielson2007interaction, teich2014systematic}, THz fields interacting with monolayer TMDs cannot excite intraexcitonic transitions, as the latter lie at considerably higher energies \cite{cha20161s}. Conceptually, THz excitation bridges the gap between dc electroabsorption (so far investigated only in the transverse configuration accessible in field effect transistors \cite{vella2017unconventional, massicotte2018dissociation}) and the coherent phenomena mentioned above. Unraveling the details of this light-matter interaction is of crucial importance for both fundamental and applied science. On a fundamental side, it would elucidate the dominant THz-induced optical nonlinearities governing the response of two-dimensional excitons with an intermediate character between the Wannier-Mott and Frenkel limit, an aspect that has remained elusive so far \cite{ogawa2010room}. On a technological side, it would establish if TMDs offer promising performances in ultrafast electro-optic devices.

\begin{figure*}[t]
	\begin{center}
		\includegraphics[width=\columnwidth]{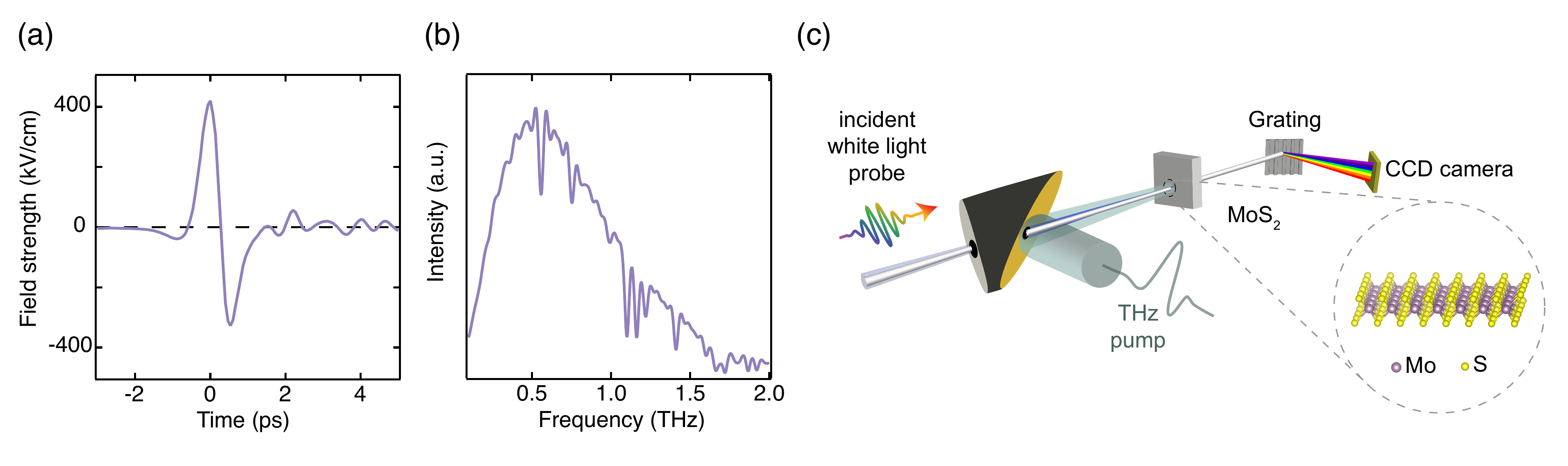}
		\caption{(a) Free-space single-cycle THz pulse generated through optical rectification of an intense 1.55 eV laser beam in a Mg:LiNbO$_3$ crystal. The maximum electric field strength is 420 kV/cm. (b) Spectrum  of the single-cycle THz field, showing a bandwidth of $\sim$ 1 THz around the central frequency of 0.54 THz. (c) Schematic illustration of our transient absorption experiment on monolayer MoS$_2$. The single-cycle THz pump and the white-light continuum probe pulses are focused on the sample. The transmitted white-light probe is spectrally dispersed inside a spectrometer and detected by a CCD camera.}
		\label{fig:Fig1}
	\end{center}
\end{figure*}

Here we study the effect of intense, single-cycle THz fields on the exciton optical properties in monolayer semiconducting TMDs. Driving these systems out of equilibrium with a strong in-plane-polarized THz pulse results in a dramatic modulation of the excitonic resonances, which instantaneously follows the THz field and is extinguished once the driving field is turned off. This light-matter interaction manifests itself through a pronounced broadening of the excitonic lineshapes, yet with an incomplete suppression of the optical resonances. These results are rationalized within the framework of Franz-Keldysh theory \cite{franz1958einfluss, keldysh1958effect}, in a regime preceding complete exciton ionization. Further estimates of the electroabsorption modulation depth involved in this process indicate superior performance of monolayer TMDs for near-infrared-to-visible electroabsorption applications, paving the way to the use of these materials as high-speed (i.e., THz) modulators and switches.

\begin{figure}[t]
	\begin{center}
		\includegraphics[width=0.7\columnwidth]{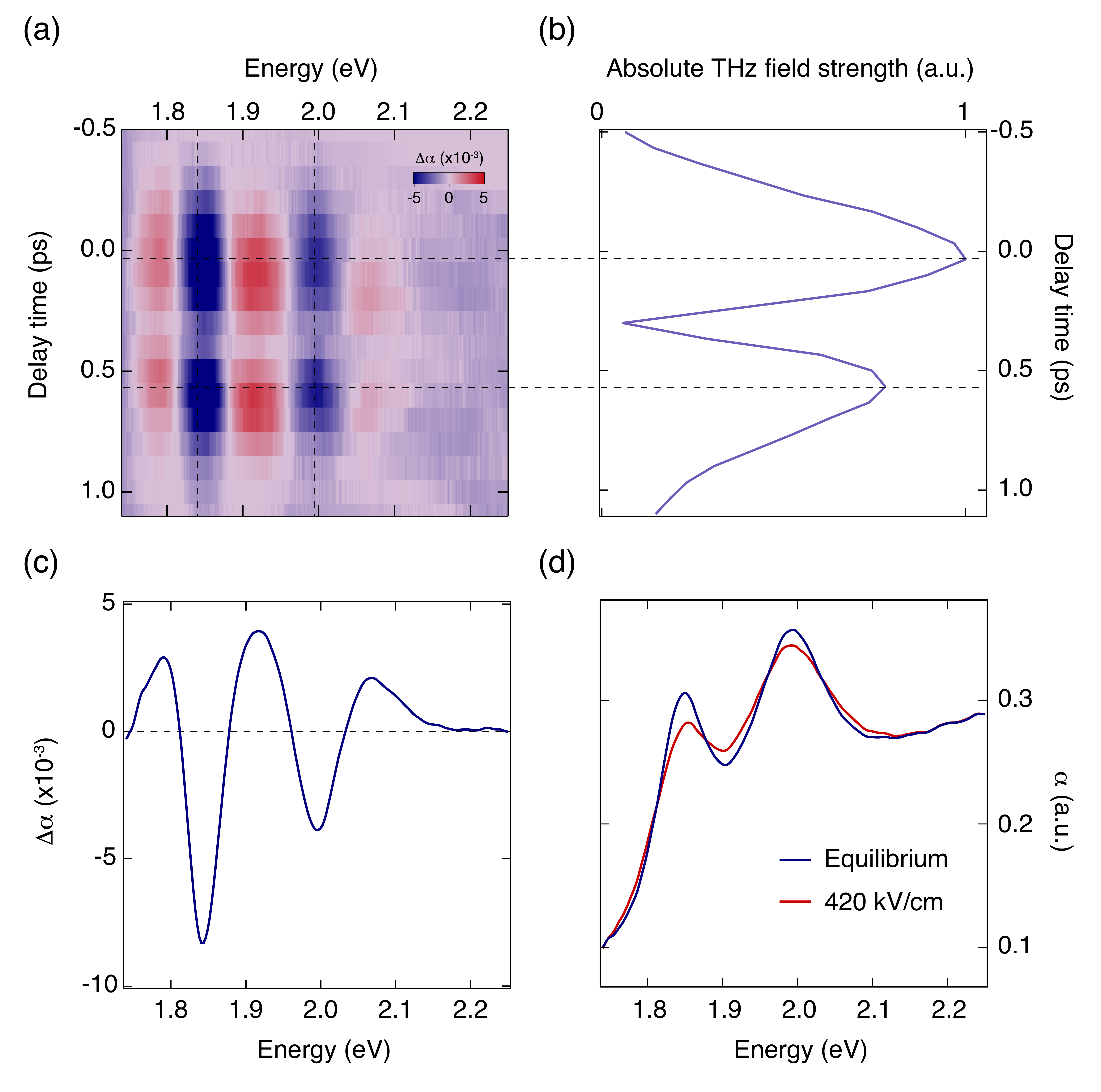}
		\caption{(a) Color-coded map of the differential absorption ($\Delta\alpha$) as a function of the probe photon energy and the time delay between pump and probe. (b) Temporal trace of the THz pulse absolute electric field strength. (c) Differential absorption spectrum at zero pump-probe delay. (d) Absorption spectrum at zero pump-probe delay plotted in equilibrium (blue trace) and upon excitation with a 420 kV/cm THz field strength (red curve). Both A and B exciton absorption profiles experience a clear broadening in the presence of the THz electric field.}
		\label{fig:Fig2}
	\end{center}
\end{figure}

As a prototypical TMD, we employ MoS$_2$ grown by chemical vapor deposition on a 0.5-mm-thick sapphire substrate. To verify its monolayer nature, the sample is characterized by spontaneous Raman scattering. Optical spectroscopy reveals that the absorption spectrum is dominated by two resonances at 1.85 and 1.99 eV, in agreement with previous studies. These peaks are the bound A and B excitons, which are made up of transitions between the spin-orbit split valence band and the lowest conduction band at the K and K$^\prime$ valleys of the Brillouin zone \cite{qiu2013optical}. When MoS$_2$ is grown on sapphire, the binding energy of these excitons is $\sim$ 240 meV \cite{park2018direct}. Their broadened lineshape in equilibrium is asymmetric between peak A and B, its origin lying in the different coherence lifetimes of the excitons due to scattering events (e.g., with phonons and impurities) of the valence band hole. More details about the sample can be found in the Supporting Information (SI).

In our experiments, we used an amplified Ti:Sapphire laser system operating at 1 kHz repetition rate with a pulse energy of 4 mJ, a central photon energy of 1.55 eV, and a pulse duration of 100 fs. Single-cycle THz pump pulses were generated by focusing about 95\% of the output beam on a Mg:LiNbO$_3$ crystal using the tilted-pulse-front phase-velocity-matching technique \cite{yeh2007generation}. The waveform of the resulting THz electric field was measured by free-space electro-optic sampling in a GaP crystal. The maximum electric field achieved at the focal position was 420 kV/cm (Fig. 1(a)) and its spectral content was centered around 0.54 THz (Fig. 1(b)). The remaining fraction of the laser output was focused on a sapphire plate to generate a weak, broadband probe beam spanning the 1.70-2.40 eV spectral range. The THz pump and continuum probe pulses were overlapped on the sample collinearly under normal incidence and the spectrally resolved transmitted probe light was detected by a CCD camera, as shown in Fig. 1(c). A thorough description of the experimental parameters is given in the Methods section. All measurements were performed at room temperature and no signs of sample degradation were observed during the entire study.

Figure 2(a) shows a color-coded map of the THz-induced transient absorption ($\Delta\alpha$) of monolayer MoS$_2$ in the proximity of the A and B exciton resonances. The data are displayed as a function of the probe photon energy and the time delay between the THz pump and the broadband visible probe. The short duration of the probe pulse allows us to precisely monitor the THz-induced changes in the exciton optical absorption. The differential absorption comprises a series of positive and negative pockets that emerge only during the first picosecond of the temporal response and extend to a photon energy of 2.10 eV. Comparing their time evolution with the absolute value of the single-cycle THz pump field profile (shown in Fig. 2(b)) reveals that the exciton absorption follows instantaneously both the first and second lobes of the THz pulse (slight deviations stem from the imperfect correction of the white-light chirp in the map of Fig. 2(a)). This confirms that the absorption changes arise from an interaction between the THz pulse and the excitons and not due to any dissipative effects from photoexcited or field-injected carriers (Fig. S3 shows an extended temporal window). Figure 2(c) shows the transient spectrum recorded at the peak of the first lobe of the single-cycle THz field, which we define as time t = 0. As already evident in Fig. 2(a), the spectrum consists of two negative features at the A and B exciton peaks, accompanied by side lobes of increased absorption. This response clearly indicates that the main effect of the THz field on the exciton resonances is to broaden them while keeping the exciton peak energies almost unchanged. These effects are confirmed by direct inspection of the total absorption spectrum under the influence of the THz pump field (red curve, Fig. 2(d)). This trace is obtained by adding the THz-induced absorption change at zero pump-probe delay to the equilibrium spectrum (shown as a blue curve in Fig. 2(d)). In the driven system (at the highest field strength used), both the A and B excitons are broadened but the only shift observed within our energy resolution is a small exciton A blueshift of 4 meV. The absorption change at the A exciton resonance is substantial, reaching a value of almost 10\%. In contrast, the THz pump excitation does not modify the spectrum above 2.10 eV, i.e. in the absorption continuum region for MoS$_2$ grown on sapphire \cite{park2018direct}.

\begin{figure}[t]
	\begin{center}
		\includegraphics[width=0.65\columnwidth]{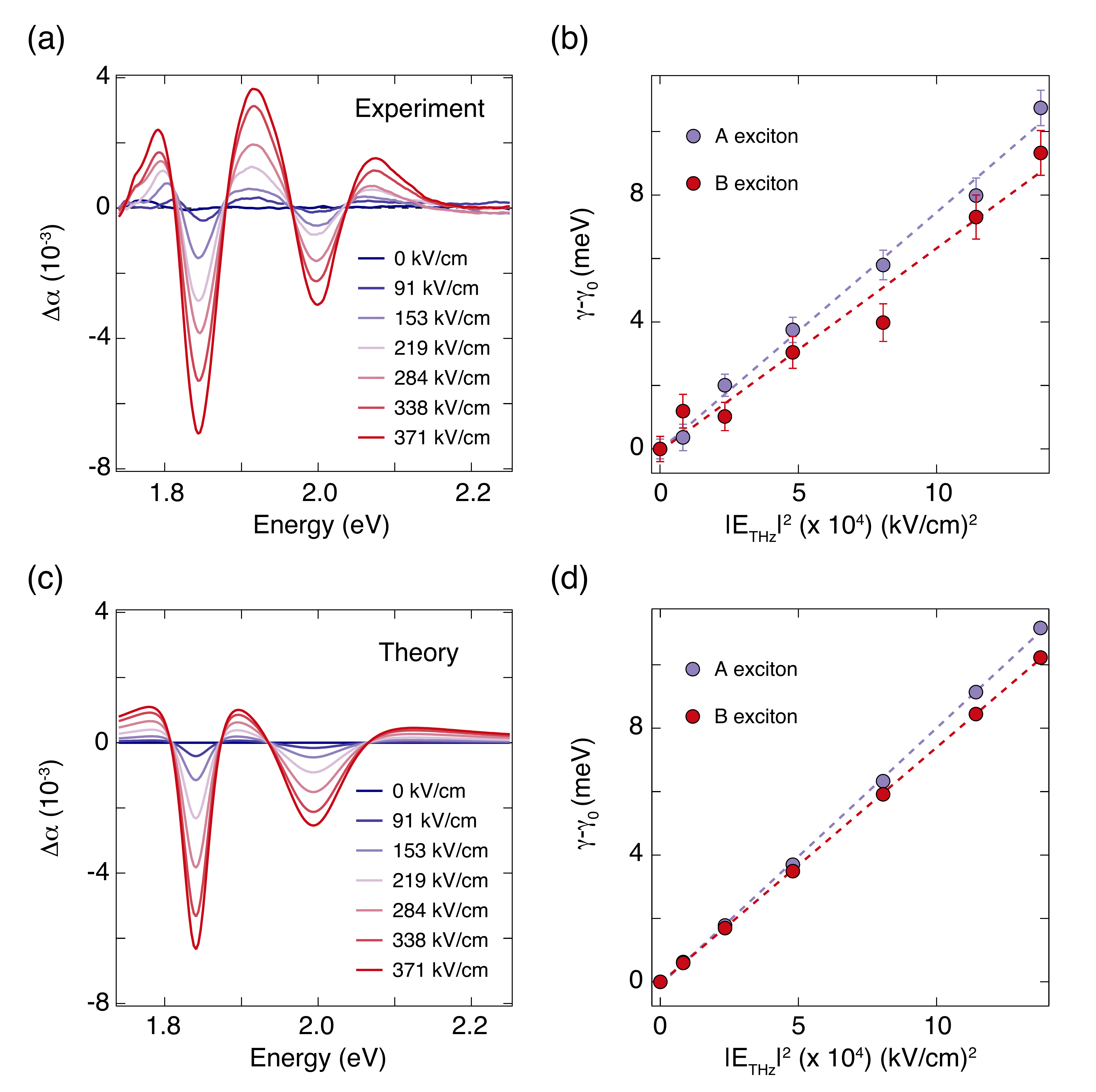}
		\caption{(a) Measured THz field dependence of the transient absorption spectrum at zero pump-probe delay. The THz field strength is varied from 0 kV/cm to 371 kV/cm. (b) Dependence of the THz-induced A and B exciton linewidths on the square of the field strength, as obtained from a Lorentz fit of the absorption spectra. Dashed lines represent linear fits to the data. $\gamma_0$ refers to the linewidth in equilibrium ($\gamma^A_0=67$~meV and $\gamma^B_0=157$~meV for the A and B exciton, respectively, as obtained from Lorentz fits of the absorption spectra in equilibrium). (c) Calculated THz field dependence of the transient absorption spectrum at zero pump-probe delay. (d) Dependence of the THz-induced A and B exciton linewidths on the square of the field strength, as obtained from the calculated spectra in panel (c). Dashed lines represent linear fits to the data. $\gamma_0$ refers to the linewidth in equilibrium.}
		\label{fig:Fig3}
	\end{center}
\end{figure}

To get more quantitative information on the redistribution of the excitonic spectral weight, we perform a systematic dependence on the THz field strength. The use of wire-grid polarizers in the pump beam path allows us to vary the THz field strength between 0 and 371 kV/cm while maintaining the THz polarization direction constant. Differential absorption spectra at zero delay time and for different THz field strengths are shown in Fig. 3(a). We observe that the signal increases gradually in intensity with no changes in shape. This suggests that the perturbative regime holds and there are no saturation effects in the explored THz field strength range. The excitons do not experience any pump-induced reduction in oscillator strength. This conclusion can be drawn by noting that the spectrally integrated value of the differential absorption (plotted in Fig. S4) remains zero at all excitation strengths. In addition, fitting the data of Fig. 3(a) with a Lorentz model (see Fig. S5 for the fit results) allows us to unveil a linear dependence of the exciton linewidth on the square of the in-plane-polarized THz field, as shown in Fig. 3(b). The degree of THz-induced broadening ($\gamma$) compared to the equilibrium intrinsic decay rate ($\gamma_0$) is similar for the A and B excitons, indicating that both excitons acquire a shorter coherence lifetime under the influence of the pump field.

A similar broadening has been reported previously in a variety of low-dimensional materials subjected to intense plane-polarized electric fields, including quantum-well structures \cite{Miller, nordstrom1998excitonic, hirori2010excitonic} and single-walled carbon nanotubes \cite{ogawa2010room}. Tunneling-ionization of excitons was proposed as a plausible underlying mechanism \cite{Dow, Miller}, but this interpretation has been contradicted by the observed scaling of the broadening on the THz field strength \cite{perebeinos2007exciton, ogawa2010room}. Furthermore, in our case, the field amplitude necessary to completely ionize the excitons in MoS$_2$ is $\sim$3 MV/cm \cite{Haastrup}, a value that is significantly higher than the 420 kV/cm used in our experiment. Therefore, to date, the microscopic origin of the exciton response to intense THz fields remains unknown.

To rationalize the mechanism behind the electroabsorption of monolayer TMDs, in the following we develop a theory based on the Franz-Keldysh effect \cite{franz1958einfluss, keldysh1958effect}. The typical manifestation of this phenomenon in direct-gap semiconductors is the formation of a finite in-gap density of states that decays exponentially from the band edge. A direct modification of the excitonic resonances in MoS$_2$ due to the presence of such a tail can be readily ruled out as it would result in a characteristic asymmetric shape of the transient spectrum. Indeed, it would lead to an asymmetric broadening of the A and B exciton resonances (due to their distinct energies) and result in a redistribution of spectral weight involving the continuum states, which is in contrast to the results shown in Fig. 3(a,b). On the other hand, an indirect manifestation of the Franz-Keldysh effect is to modify the continuum states and enhance the scattering rate for the excitons into these states. This process is more favorable if the energy difference between the exciton and the final continuum state is smaller, which is the case for the in-gap states created by the Franz-Keldysh effect \cite{jauho1996dynamical}. This leads to a change in the environment-mediated (e.g., thermally excited phonons) exciton scattering, which in turn is reflected in a symmetrical broadening of the excitonic peaks and a local spectral weight redistribution (i.e. in the energy region surrounding each exciton).

To describe the effect of the field on the broadening of the exciton resonances, we rely on the Redfield equation \cite{Dow}. We assume the process to be adiabatic and therefore describable with a static electric field. In the model, the exciton decay is represented by the transition from the bound excitonic state to continuum states due to the interaction with the environment, which is modeled by the Ohmic bath with the characteristic cutoff frequency $\Omega_c$ (set at 49~meV/$\hbar$ for exciton A and 66~meV/$\hbar$ for exciton B for the best fits to experimental results). We remark that the choice of a different bath would not modify our conclusions. Once the strong field is applied to the system, the density of states of the final continuum state is renormalized by the Franz-Keldysh effect, and the scattering channels are modified. As a result, the exciton decay rate is enhanced by the field as $\gamma$ = $\gamma_0$  exp$(E_{THz}^2/24\mu \Omega_c^3)\approx \gamma_0 (1+E_{THz}^2 /24\mu\Omega_c^3)$, where $E_{THz}$ is the applied electric field strength and $\mu=0.24 m_e$ is the electron-hole reduced mass\cite{reducedmass}(where $m_e$ refers to the electron mass). The complete derivation is presented in $\S$S2. The change of the scattering rate is directly reflected in the broadening of the exciton resonance. Figure 3(c) shows the theoretical differential absorption change under the effect of a static field of different strengths, computed for the two main excitonic peaks of MoS$_2$. Comparing Figs. 3(a) and 3(c) confirms that the Franz-Keldysh mechanism captures the salient shapes of the experimentally observed transient absorption spectra. The deviations between the experiment and the theory are most likely caused by slight asymmetries of the exciton lineshapes in the measured spectra (which are modelled with simple Lorentzian functions). The quadratic field strength dependence of the Franz-Keldysh mechanism yields the calculated exciton linewidths shown in Fig. 3(d), to be compared with Fig. 3(b). To our knowledge, this is the first time that the origin of the THz-induced exciton nonlinearities can be fully described theoretically. As such, our results assume a particular importance as they can predict the exciton behavior in a wide variety of low-dimensional systems subject to intense THz radiation.

Finally, we discuss the impact of our findings on possible electro-optic applications. The important upshot of our results is that monolayer MoS$_2$ films provide a large modulation depth around 0.05 dB/nm at room temperature when exposed to 420 kV/cm THz fields. This large tunability is ascribed to the combination of in-plane electric field and strong exciton electro-optical nonlinearities in monoalyer MoS$_2$ \cite{pedersen2016exciton}. A simple estimate that relies on the dielectric function of the material yields a THz-induced variation in the linear refractive index around 2.2\%, which indicates that the phase modulation provided by a single layer of MoS$_2$ is around an order of magnitude higher than traditional ferroelectric electro-optic materials (see Fig. S8 for the detailed comparison). The remarkably high modulation depth achieved in a single monolayer opens intriguing perspectives towards the use of these semiconductors (stacked as heterostructures) as compact and efficient electroabsorption modulators for integrated photonics. 

In conclusion, we demonstrated that the application of intense THz fields to monolayer TMDs produces a substantial broadening of their exciton resonances, leading to a giant electroabsorption response that instantaneously follows the THz field. Future extensions of these studies to circularly-polarized THz fields will offer valley-selectivity for this coherent light-matter interaction; the use of even higher field strengths will promote complete, dynamic dissociation of excitons into free electron-hole pairs, a feature that is desirable in many optoelectronic applications.

\clearpage
\newpage

\section{Methods}

\noindent \textit{Sample synthesis and characterization}\\
Monolayer MoS$_2$ was grown by the chemical vapor deposition (CVD) method on a 0.5 mm thick sapphire substrate. Prior to the growth, the sapphire substrate was cleaned by deionized water, acetone, and isopropyl alcohol sequentially. Perylene-3,4,9,10-tetracarboxylic potassium salt (PTAS) molecules were used as the seeding promoter and were coated onto two clean SiO$_2$/Si pieces, which serve as the seed reservoirs to provide sufficient seeding molecules for MoS$_2$ synthesis during the growth. The target sapphire substrate was suspended between those two PTAS-coated SiO$_2$/Si seed reservoirs. All of these three substrates were faced down and placed on a crucible containing molybdenum oxide (MoO$_3$, 99.98\%) powder precursor. This MoO$_3$ precursor was put in the middle of a 1 inch quartz tube reaction chamber and another sulfur powder (99.98\%) precursor was placed upstream, 14 cm away from the MoO$_3$ crucible, in the quartz tube. Before heating, the CVD system was purged using 1000 standard cubic centimeters per minute (sccm) of Ar (99.999\% purity) for 5 min, and then 20 sccm of Ar was introduced into the system as a carrier gas. Next, the temperature of the reaction chamber was increased to 625 $^{\circ}$C at a rate of 30 $^{\circ}$C min$^{-1}$. The monolayer MoS$_2$ was synthesized at 625 $^{\circ}$C for 3 min under atmospheric pressure. The temperature at the position where the sulfur was located was around 180 $^{\circ}$C during growth. Finally, the system was cooled down to room temperature quickly. During the cooling process, 1000 sccm Ar flow was employed into the chamber to remove the reactants, preventing further unintentional reactions. Imaging the sample with atomic force microscopy allowed us to estimate the domain size around 5-10 $\mu$m on average (Fig. S1). The monolayer character of the sample was confirmed via spontaneous Raman scattering. The Raman spectrum of Fig. S2(a) shows that the difference between the peak energy of the $E_{2g}^1$ mode and that of the $A_{1g}$ mode is 2.41 meV. This indicates that the sample is mostly monolayer \cite{chakraborty2013layer}. The small shoulder around 51.6 meV is due to the $A_{1g}$ mode of the sapphire substrate. The sample quality of monolayer MoS$_2$ was confirmed by optical absorption spectroscopy. Two absorption peaks at 1.85 eV and 1.99 eV in Fig. S2(b) correspond to the A and B exciton transitions respectively, which agrees with the results previously reported in the literature\cite{li2014measurement}. \\

\noindent \textit{Experimental set-up}\\
High-field THz pulses were generated in Mg:LiNbO$_3$ crystal by tilting the pulse front to achieve phase matching \cite{yeh2007generation}. By using a three-parabolic-mirror THz imaging system, the image of the THz spot on the sample was confined close to its diffraction limit. The incident THz field was measured in the time domain using electro-optic sampling with a 100 $\mu$m-thick (110)-oriented GaP crystal. When pumping with a 4 W laser from an amplified Ti:Sapphire laser system (repetition rate 1 kHz, central photon energy 1.55 eV, pulse duration 100 fs), the maximum electric field of the THz pulses reaches 420 kV/cm at the focus, with a spectrum centered at 0.54 THz. This beam was used as the pump arm. Using wire-grid polarizers, we were able to perform a THz field-strength dependence while maintaining the THz polarization direction constant. For the probe arm, 5\% of the full pulse energy was focused onto a sapphire plate to generate a white-light continuum via self-phase modulation. The THz pump pulse and white-light probe pulse (spannning the 1.70-2.40 spectral range) are overlapped on the sample collinearly under normal incidence. The sample image was magnified by a factor of 2 using a 4-f lens system onto the spectrometer slit. The transmitted white light was spectrally resolved by a spectrometer (Andor Shamrock) and detected by a CCD camera (Newton 920). To obtain the change in transmittance ($\Delta$T/T) induced by the THz pulse, we synchronously chopped the THz pump beam at 0.5 kHz, thereby blocking every other pump beam. The absorption spectrum was also measured in the same set-up by simply switching the femtosecond white-light source to a halogen lamp source. By marking the position of the THz and white light overlapping spots in the spectrometer, the halogen lamp light was also focused on the same spot to ensure that all the data were collected at the same sample spot with an approximate accuracy of 10 $\mu$m. All the optical measurements were performed at ambient conditions. The acquired data were subsequently corrected to account for the white-light chirp.

\clearpage
\newpage

\begin{acknowledgement}
$\bigtriangleup$ These authors contributed equally to this work. We acknowledge helpful discussions with J. Yoon, Y. Bie, E. J. Sie, B. Skinner, P. Sivarajah and technical assistance from T. McClure. This manuscript is based upon work by J.S. and K.A.N. supported in part by the U.S. Army Research Laboratory (ARL) and the U.S. Army Research Office through the Institute for Soldier Nanotechnologies, under Cooperative Agreement number W911-NF-18-2-0048. J.S. and K.A.N. acknowledge additional support from the Samsung Global Outreach Program. E.B. acknowledges additional support from the Swiss National Science Foundation under fellowships P2ELP2-172290 and P400P2-183842. N.G. and E.B. acknowledge support from DOE, BES DMSE. S.L. acknowledges support from the Alexander von Humboldt foundation. S.A.S. acknowledges JST-CREST under Grant No. JP-MJCR16N5. This work was supported by the European Research Council (ERC-2015-AdG694097), the Cluster of Excellence  (AIM), Grupos Consolidados (IT1249-19) and SFB925. The Flatiron Institute is a division of the Simons Foundation. P.C.S. and J.K. acknowledge the financial support from the Center for Energy Efficient Electronics Science (NSF Award No. 0939514).
\end{acknowledgement}


\newpage
\clearpage


\renewcommand{\thesection}{S\arabic{section}}  
\renewcommand{\thetable}{S\arabic{table}}  
\renewcommand{\thefigure}{S\arabic{figure}} 
\setcounter{figure}{0}

\section{S1. Evaluation of the transient absorption}

To evaluate the transient absorption ($\Delta\alpha$) from the measured differential transmittance ($\Delta$T/T), we applied a common approach known in the literature \cite{sie2015intervalley}. Specifically, we performed a Kramers-Kronig constrained variational analysis \cite{kuzmenko2005kramers} to extract the complex dielectric function ($\epsilon$ = $\epsilon_1$ + i$\epsilon_2$) from the static transmittance T. The results we obtain are consistent with those reported in previous studies \cite{ref:yi}. We then used the dielectric function to obtain the absorbance $\alpha$, as reported previously \cite{sie2015intervalley}. We applied this procedure for all the acquired data at different THz pump field strengths. Finally, at each field strength, we performed a Lorentz analysis of the absorption spectra in order to estimate the parameters for the exciton resonances. The results of the fits are superposed as solid lines over the experimental data in Fig. S5. While the oscillator strength and the exciton peak energy are essentially unchanged under THz excitation, the resonance linewidth undergoes a significant modification with increasing field strength, as shown in Fig. 3(b).

\section{S2. Theoretical modeling with the Franz-Keldysh effect}

The time-resolved results show that the field-induced broadening of the excitonic peaks occurs only during the presence of the THz field. This observation points towards an adiabatic effect which could be interpreted with a static field. Our proposal is that the static Franz-Keldysh effect plays a relevant but indirect role. The textbook manifestation of the Franz-Keldysh effect is the formation of a finite in-gap density of states with an exponential tail from the electronic bandgap. If such a tail extended far enough to shift and broaden the excitonic resonances, it would result in an asymmetric shape of the A and B excitonic resonances (as the energies of the two resonances are different). Another possibility is that the electric field of the THz pulse induces direct dissociation of the excitons, consequently reducing their lifetime and broadening their lineshapes. However, according to previous studies \cite{haastrup2016, massicotte2018}, the required field magnitude for direct dissociation should be much higher than those utilized in our experiments and the scaling would be different than that observed in our data.

Here, we propose an indirect role of the Franz-Keldysh effect through the modification of particle-hole continuum. Specifically, the Franz-Keldysh effect provides a larger number of particle-hole states below the electronic gap, thus opening additional scattering channels for the excitons and increasing the broadening of their resonances. The scattering of the exciton into a continuum state has to be mediated by the environment (e.g., thermal phonons) and is more favorable if the energy difference between the exciton and the final continuum state is smaller, which is the case for the in-gap states created by the Franz-Keldysh effect.

In the following paragraphs, we analyze the above scenario with a simple model based on the Franz-Keldysh effect and the Redfield equation.

%===============================================================================
\subsection{Franz-Keldysh effect in 2D systems \label{sec:2d-FKE}}
%===============================================================================

Here, we derive the particle-hole continuum density of states, also referred to as the joint density of states, of a two-dimensional material (2D) under the influence of a static electric field. This derivation represents the extension of the Franz-Keldysh effect to the 2D case. Let us start from the time-dependent expression for the density of states as in Eqs.~(2) and (3) of Jauho and Johnsen \cite{PhysRevLett.76.4576}.
\be
\rho(\omega, T) = \frac{1}{2\pi} \sum_{\vecb k}\tilde A(\vecb k, \omega, T),
\label{eq:jdos-jauho}
\ee
\be
\tilde A(\vecb k,\omega, T) = \int d\vecb r d\tau e^{iw} \int d\vecb p
\times \exp \left \{-i \int^{T+\tau/2}_{T-\tau/2} dt_1 \epsilon \left[
\vecb p - \vecb A(t_1)
\right] \right \},
\label{eq:spectral-jauho}
\ee
where $\vecb A(t)$ is the vector potential, and $w$ is defined as
\be
w\equiv \tau \omega -\vecb r \cdot \vecb k - \int^{T+\tau/2}_{T-\tau/2}
\frac{dt_1}{\tau} \vecb r \cdot \vecb A(t_1).
\label{eq:def-w}
\ee

In the following, we consider only a static electric field $\vecb E_0$, which can be described by the following vector potential
\be
\vecb A(t) = - \vecb E_0 \times (t-T),
\ee
which, according to Eq.~(\ref{eq:def-w}) gives
\be
w\equiv \tau \omega -\vecb r \cdot \vecb k,
\ee
and Eq.~(\ref{eq:spectral-jauho}) can be rewritten as
\be
\tilde A(\vecb k, \omega) = \frac{1}{(2\pi)^2}
\int^{\infty}_{-\infty}  d\tau e^{i\omega \tau}
\times \exp \left \{-i \int^{T+\tau/2}_{T-\tau/2} dt_1 \epsilon \left[
\vecb k - \vecb A(t_1)
\right] \right \}.
\label{eq:spectral-jauho-2}
\ee

We assume a parabolic band dispersion for particle-hole pairs, which reads
\be
\epsilon(\vecb k) = \epsilon_g + \frac{1}{2\mu} \vecb k^2,
\ee
where $\epsilon_g$ is the electronic band gap of the crystal, and $\mu$ is the reduced electron-hole mass.
Under this assumption, the dynamical phase factor in Eq.~(\ref{eq:spectral-jauho-2})  can be evaluated as
\be
\int^{T+\tau/2}_{T-\tau/2} dt_1 \epsilon \left[
\vecb k - \vecb A(t_1)
\right] 
= \int^{\tau/2}_{-\tau/2} dt_1 \epsilon \left[
\epsilon_g + \frac{1}{2\mu} \left (\vecb k + \vecb E_0 \tau \right )^2
\right]
= \left (\epsilon_g + \frac{\vecb k^2}{2\mu} \right )\tau
+ \frac{\vecb E^2_0}{8\mu} \frac{\tau^3}{3}
\label{eq:dynamical-phase-factor}.
\ee
Inserting Eq.~(\ref{eq:dynamical-phase-factor}) into Eq.~(\ref{eq:spectral-jauho-2}),
we have:
\be
\tilde A(\vecb k, \omega) = \frac{1}{(2\pi)^d}
\int^{\infty}_{-\infty}  d\tau \exp \left \{
-i \left [ \left (\epsilon_g +\frac{\vecb k^2}{2\mu} -\omega \right )\tau
+\frac{\vecb E^2_0}{8\mu}\frac{\tau^3}{3}
\right ]
\right \}
\label{eq:spectral-jauho-3}
\ee
Employing the Airy function, $\mathrm{Ai}(x)$, defined as
\be
\mathrm{Ai}(x) = \frac{1}{2\pi} \int^{\infty}_{-\infty} e^{i(t^3/3+tx)}
= \frac{1}{2\pi} \int^{\infty}_{-\infty} \cos \left [t^3/3+tx \right ]
= \frac{1}{\pi} \int^{\infty}_{0} \cos \left [t^3/3+tx \right ],
\label{eq:airy}
\ee
the spectral function $\tilde A(\vecb k, \omega)$ in Eq.~(\ref{eq:spectral-jauho-3})
can be rewritten as
\be
\tilde A(\vecb k, \omega) = \frac{1}{(2\pi)^2} \frac{2\pi}{\beta}
\mathrm{Ai} \left (
\frac{\epsilon_g + \frac{\vecb k^2}{2\mu} -\omega}{\beta}
\right ),
\label{eq:spectral-jauho-4}
\ee
where we introduced $\beta$ as
\be
\beta \equiv \left (\frac{\vecb E^2_0}{8\mu} \right )^{1/3}.
\ee
The spectral function we just obtained corresponds to Eq.~(7) of Ref.~\cite{PhysRevLett.76.4576}.

We proceed with the evaluation of the joint density of states by inserting Eq.~(\ref{eq:spectral-jauho-4}) into Eq.~(\ref{eq:jdos-jauho}),
\be
\rho^{2D}(\omega) &=&\frac{1}{2\pi}\int d\vecb k \frac{1}{(2\pi)^2}\frac{2\pi}{\beta}
\mathrm{Ai} \left (
\frac{\epsilon_g + \frac{\vecb k^2}{2\mu} -\omega}{\beta}
\right ) \nonumber \\
&=& \frac{1}{2\pi}\frac{1}{\beta} \int^{\infty}_0 dk \cdot k \cdot
\mathrm{Ai} \left (
\frac{\epsilon_g + \frac{k^2}{2\mu} -\omega}{\beta}
\right ).
\label{eq:jdos-2d-1}
\ee
Applying the variable transformation, $x=k^2/2\mu \beta$, Eq.~(\ref{eq:jdos-2d-1}) can be rewritten as
\be
\rho^{2D}(\omega) &=& \frac{\mu}{2\pi} \int^{\infty}_0 dx 
\mathrm{Ai} \left (
\frac{\epsilon_g -\omega}{\beta} + x
\right ) \nonumber \\
&=& \frac{\mu}{2\pi} \int^{\infty}_{-\frac{\omega-\epsilon_g}{\beta}} dy
\mathrm{Ai} (y),
\label{eq:jdos-2d}
\ee
where we further transformed the integration variable $y=x+(\epsilon_g -\omega)/\beta$.

We can readily see that, in the weak field limit, $\beta \rightarrow 0$ $(E_0\rightarrow 0)$, the joint density of state expression above becomes a Heaviside step function that is typical of 2D systems
\be
\rho^{2D}(\omega) \approx \frac{\mu}{2\pi}  \Theta \left ( \omega-\epsilon_g\right ).
\ee

Figure S6 shows how the 2D joint density of states is affected by the external electric field with different magnitudes. We can distinguish a typical exponential tail for energies below and an oscillatory behavior above the electronic gap, similar to the three-dimensional case.
%

%
%===============================================================================
\subsection{Theoretical analysis of field enhanced decay of excitons \label{sec:bath-decay}}
%===============================================================================

In this paragraph, we derive an expression for the field-induced enhancement of the exciton decay by employing the modified density of states in Eq.~(\ref{eq:jdos-2d}). We consider a system that consists of a sub-system and a bath described by the Hamiltonian
\be
H_{Tot} = H_S + H_B + H_{SB},
\ee
where $H_S$ is the subsystem Hamiltonian, $H_B$ is the bath Hamiltonian, and $H_{SB}$ is the coupling between the subsystem and the bath.
Furthermore, we assume that the bath consists of a series of harmonic oscillators as
\be
H_B = \sum_a \left [\frac{\hat P^2_a}{2M_a} + \frac{1}{2}M_a\Omega_a \hat R^2_a  \right ],
\ee
where $M_a$ is mass of a harmonic oscillator, $\Omega_a$ is its eigenfrequency, $\hat P_a$ is its momentum operator and $\hat R_a$ is its position operator.
In this work, we consider the following linear coupling form for the coupling Hamiltonian $H_{SB}$ as
\be
H_{SB} = g\sum_a \hat A \otimes \hat R_a,
\ee
where $g$ is a coupling constant and $\hat A$ is a sub-system operator.

According to Ref.~\cite{2019arXiv190200967L}, under the Born and Markov approximations, one can derive the Redfield equation in the interaction picture,
\be
\frac{d\rho(t)}{dt} = -g^2 \sum_a \int^t_0 d\tau \left \{
\mathcal{B}_{aa}(\tau) 
\left [A(t), A(t-\tau) \rho(t) \right ] +\mathrm{h.c.}
\right \},
\label{eq:redfield}
\ee
with $\mathcal{B}_{aa}(\tau)$ is the correlation function of the $a$ harmonic oscillator, which reads
\be
\mathcal{B}_{aa}(\tau) = \mathrm{Tr}\left \{
\hat R_a(0) \hat R_a(-\tau)  \rho_B
\right \}
= \frac{1}{2M_a\Omega_a} \left [e^{-i\Omega_a\tau}+\frac{2}{e^{\beta\hbar\Omega_a}-1}
\cos \Omega_a\tau \right ],
\label{eq:ho-correlation-function}
\ee
where $\rho_B$ is the bath density matrix in thermal equilibrium, and $\beta$ corresponds to the inverse temperature, $\beta \equiv 1/k_BT$. Then, we introduce the total correlation function of the bath as
\be
\mathcal{B}(\tau) = g^2\sum_a \mathcal{B}_{aa}(\tau) 
= \int^{\infty}_0 d\Omega 
J(\Omega) \left [e^{-i\Omega\tau}+\frac{2}{e^{\beta\hbar\Omega}-1}
\cos(\Omega\tau) \right ],
\label{eq:bath-correlation-function}
\ee
where the bath spectral density $J(\Omega)$ is given by
\be
J(\Omega) \equiv \sum_a \delta(\Omega-\Omega_a)\frac{g^2}{2M_a\Omega_a}.
\ee

Employing this bath correlation function, Eq.~(\ref{eq:bath-correlation-function}) can be rewritten as
\be
\frac{d\rho(t)}{dt} = - \int^t_0 d\tau 
\mathcal{B}(\tau) 
\left [A(t), A(t-\tau) \rho(t) \right ] +\mathrm{h.c.}
\label{eq:redfield-2}
\ee

We then proceed with the evaluation of the population transfer rate from the excitonic state $|e\rangle$ to a final state $|f\rangle$ (a continuum particle-hole state) via the interaction with the bath as
\be
\gamma_f &=& \frac{\langle f|\rho(T)|f \rangle - \langle f|\rho(0)|f \rangle}{T}
\nonumber \\
&=&-\frac{1}{T} \mathrm{Tr} \left \{|f\rangle \langle f|
\int^T_0 dt \int^t_0 d\tau 
\mathcal{B}(\tau) \left \{
\left [A(t), A(t-\tau) \rho(t) \right ] +\mathrm{h.c.} \right \}\right \}
\nonumber \\
&\approx&-\frac{1}{T} \mathrm{Tr} \left \{|f\rangle \langle f|
\int^T_0 dt \int^\infty_0 d\tau 
\mathcal{B}(\tau) \left \{
\left [A(t), A(t-\tau) \rho(t) \right ] +\mathrm{h.c.} \right \}\right \},
\ee
where, in the last step, we changed integration limits assuming that the period $T$ is much longer than the decay time of the correlation function $\mathcal{B}(\tau)$. We further assume that the system is initially in the exciton state, $\rho(0)=|e\rangle\langle e|$ and that the subsystem-bath coupling is weak. As a result, one can evaluate the leading term of the scattering rate as
\be
\gamma_f &\approx& -\frac{1}{T}\mathrm{Tr} \left \{|f\rangle \langle f|
\int^T_0 dt \int^\infty_0 d\tau 
\mathcal{B}(\tau) \left \{
\left [A(t), A(t-\tau) |e\rangle\langle e| \right ] +\mathrm{h.c.} \right \}\right \}
\nonumber \\
&=&\int^\infty_0 d\tau 
\mathcal{B}(\tau) 
|\langle f|A|e\rangle |^2 e^{ -i\left (\epsilon_f -\epsilon_e \right )\tau } + c.c.,
\label{eq:decay-rate-gf}
\ee
where $\epsilon_e$ and $\epsilon_f$ are the energy of the exciton and the final state, respectively. In the last line of Eq.~(\ref{eq:decay-rate-gf}), we used the definition of the operator $A(t)\equiv e^{iH_St}Ae^{-iH_St}$. Inserting Eq.~(\ref{eq:bath-correlation-function}) into Eq.~(\ref{eq:decay-rate-gf}), the following expression is obtained:
\be
\gamma_f = 2\pi |\langle f|A|e\rangle|^2 J(\Omega) \frac{1}{e^{\beta \hbar\Omega}-1} \Bigg |_{\hbar\Omega=\epsilon_f-\epsilon_e},
\ee
where $\epsilon_f-\epsilon_e>0$ is assumed.

By adding contributions from all possible final states, the decay rate of the exciton via the environment can be evaluated as $\gamma = \sum_f \gamma_f $. Further assuming that matrix elements $|\langle f|A|e\rangle|^2$ do not depend on the final states, $|\langle f|A|e\rangle|= \mathcal M^2$, one can evaluate the decay rate as
\be
\gamma = \sum_f 2\pi \mathcal M^2 J(\Omega) \frac{1}{e^{\beta \hbar\Omega}-1} \Bigg |_{\hbar\Omega=\epsilon_f-\epsilon_e} 
=2\pi \mathcal M^2 \int^{\infty}_0 d\Omega J(\Omega) n_B(\Omega) \rho_D(\Omega),
\label{eq:decay-rate}
\ee
where $n_B(\Omega)$ is the Bose-Einstein distribution, $n_B(\Omega)=(e^{\beta\hbar\Omega}-1)^{-1}$, and 
the density of states of the final system $\rho_{D}(\omega)$ is introduced as
\be
\rho_{D}(\omega) = \sum_f \delta(\epsilon_f -\epsilon_e - \omega ).
\ee

A standard assumption for harmonic oscillator bath is to assume an Ohmic spectral density,
\be
J(\Omega) = \eta \Omega e^{-\Omega/\Omega_c},
\ee
where $\eta$ is a coupling strength parameter, and $\Omega_c$ is the cutoff frequency. 
Further assuming the high-temperature limit, $\hbar \Omega_c/k_BT \ll 1$, for the Bose-Einsten distribution, the decay rate
in Eq.~(\ref{eq:decay-rate}) is described as
\be
\gamma = 2\pi \eta k_BT \mathcal M^2 \int^{\infty}_0 d\omega 
\exp\left [
-\frac{\omega}{\Omega_c}
\right ] \rho_{D} (\omega).
\label{eq:decay-rate-high-tmp}
\ee

Employing Eq.~(\ref{eq:jdos-2d}) as the density of states of 2D materials under an electric field, the decay rate $\gamma$ can be evaluated as
\be
\gamma =  \frac{\gamma_0}{\Omega_c e^{-\epsilon_e/\Omega_c}}
\frac{2\pi}{\mu}
\int^\infty_0 d\omega e^{-\frac{\omega}{\Omega_c}} \rho^{2D}(\omega),
\label{eq:decay-rate-2d}
\ee
where $\gamma_0$ is the intrinsic decay rate without the applied electric field, which is recovered in the limit of $E_0 \rightarrow 0$. 

Furthermore, if the external field is not strong enough to induce direct ionization, the decay rate of Eq.~(\ref{eq:decay-rate-2d}) can be approximated as
\be
\gamma &\approx&
\frac{\gamma_0}{\Omega_c e^{-\epsilon_e/\Omega_c}}
\frac{2\pi}{\mu}
\int^\infty_{-\infty} d\omega e^{-\frac{\omega}{\Omega_c}} \rho^{2D}(\omega)
= \gamma_0 \exp \left [ \frac{\vecb E^2_0}{8\mu}\frac{1}{3\Omega^3_c} \right ].
\label{eq:decay-rate-approx}
\ee

From this expression we can explicitly evaluate the effect of the external electric field on the excitonic linewidth.

Finally, from Eq.~(\ref{eq:decay-rate-approx}), one can clearly see that in the weak field limit  the enhancement of the decay rate is proportional to the square of the electric field strength,
\be
\gamma = \gamma_0 \exp \left [ \frac{\vecb E^2_0}{8\mu}\frac{1}{3\Omega^3_c} \right ]
\approx \gamma_0 \left [1 + \frac{\vecb E^2_0}{8\mu}\frac{1}{3\Omega^3_c} \right  ],
\label{eq:decay-rate-approx-2}
\ee
This finding is consistent with the experimental evidence that the broadening is linear in the field intensity as shown in the main text.

To evaluate the transient absorption and the exciton linewidth in Figs.~3(c,d) in the main text with Eq.~(\ref{eq:decay-rate-approx-2}),
we set the intrinsic excitonic linewidth $\gamma_0$ to $\gamma^A_0=67$~meV and $\gamma^B_0=157$~meV for the A and B excitons, obtained from Lorentz fits of the absorption spectra in equilibrium.
Furthermore, for the best fits to experimental results, the cutoff frequencies $\Omega_c$ were set to $\Omega^A_c=49$~meV/$\hbar$ and $\Omega^B_c=66$~meV/$\hbar$ for the A and B excitons, respectively.
We used the common excitonic mass $\mu=0.24 m_e$ for the A and B excitons.\cite{reducedmass}

\section{S3. Evaluation of the electro-optic coefficients}

In this section, we provide an estimate of the electro-optic coefficients characterizing our monolayer MoS$_2$ film under the influence of intense THz fields. When irradiating the material with a THz field strength of 420 kV/cm, near 1.90 eV we observe that the real part of the refractive index is modulated by $\sim$2.2$\%$ (Fig. S7). From this value, we can extract the electro-optic coefficient $r_{ij}$ by using the relation $\Delta (1/n^2)_{i}=\sum\limits_{j}r_{ij}E_{j}$, where $E_{j}$ is the applied electric field and $i,j$ represents $x$, $y$ and $z$. Formally, $r_{ij}$ is a tensor. However, in our experiment, we can simplify it to a scalar value $r_{eff}$ by considering that our monolayer MoS$_2$ film has randomly oriented domains \cite{Boyd}. Since the electric field screening needs to be taken into account in the practical performance of a phase modulator, the value of the (quasi-static) dielectric permittivity $\epsilon_{DC}$ becomes an important parameter. Therefore, the performance of a material is measured through the unitless quantity $n^3r_{eff}/\epsilon_{DC}$ \cite{Boyd}. Due to the 2D confinement in the monolayer limit, the value of $\epsilon_{DC}$ in MoS$_2$ is several times larger than that of traditional ferroelectric electro-optic materials. Thus, in an exciton-based electro-optic modulator made out of monolayer MoS$_2$, the phase modulation capability is much higher than that offered by traditional materials used in electro-optical modulators. For example, evaluating $n^3r_{eff}/\epsilon_{DC}$ at the He-Ne laser wavelength of 633 nm yields a factor of 5 compared to other materials (see Fig. S8). At 650 nm, the increase would be by more than an order of magnitude ($n^3r_{eff}/\epsilon_{DC}$ = 150). The time evolution of the THz-induced modulation of the excitons in MoS$_2$ also demonstrates that an exciton-based electro-optic modulator made out of this material possesses an ultrabroad bandwidth of several THz. Similar conclusions can be drawn when considering the THz-induced modulation of the absorption coefficient, which is relevant for the development of electroabsorption modulators. In an experiment utilizing the transmission geometry like ours, the electroabsorption performance of a material subjected to in-plane electric field can be characterized by the absorption modulation depth. At a field strength of 420 kV/cm, monolayer MoS$_2$ shows switching from 89.58\% transmission to 90.48\% around 1.90 eV, which corresponds to an absorption modulation depth around 0.05 dB/nm.

\newpage

\begin{figure}[h]
	\begin{center}
		\includegraphics[width=0.6\columnwidth]{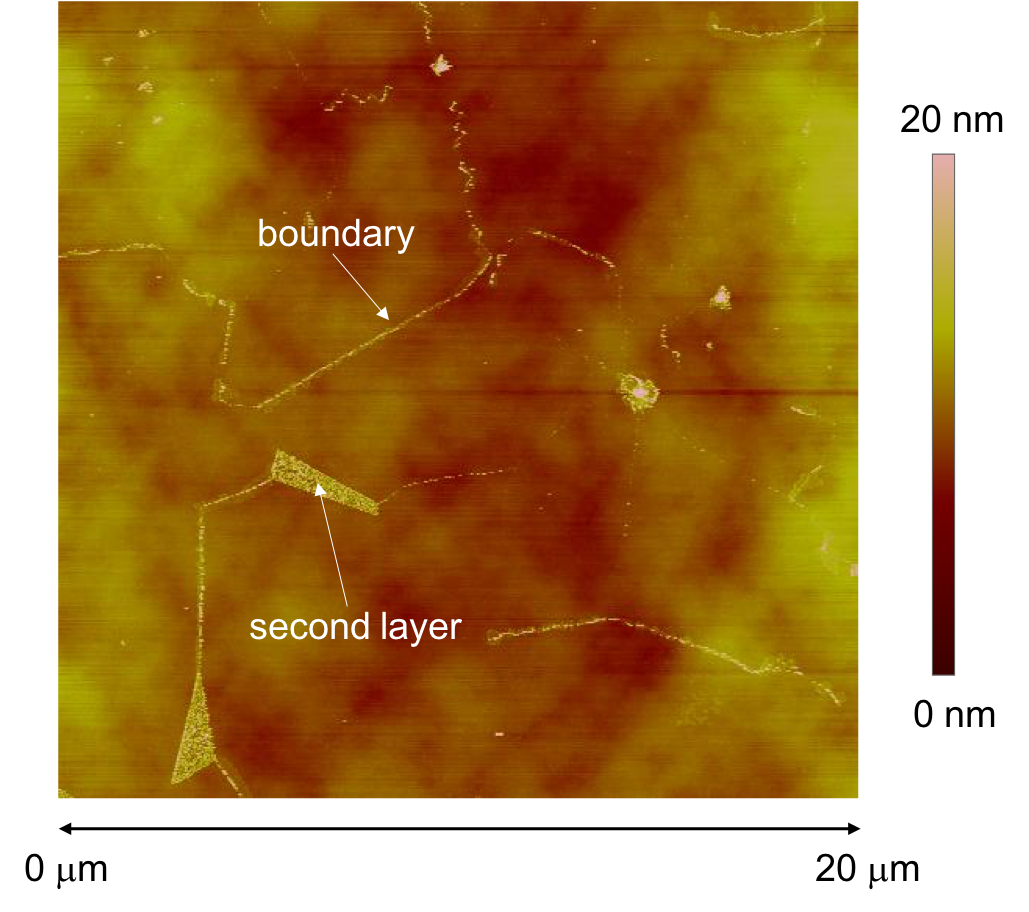}
		\caption{Atomic force microscope image of the surface of CVD-grown MoS$_2$. The imaged region is 20 $\mu$m wide and the color bar refers to the surface roughness. We  estimate the domain size in our sample to be 5-10 $\mu$m on average. A boundary between two domain and a second layer are indicated.}
		\label{fig:FigS1}
	\end{center}
\end{figure}
\newpage

\begin{figure}[h]
	\begin{center}
		\includegraphics[width=\columnwidth]{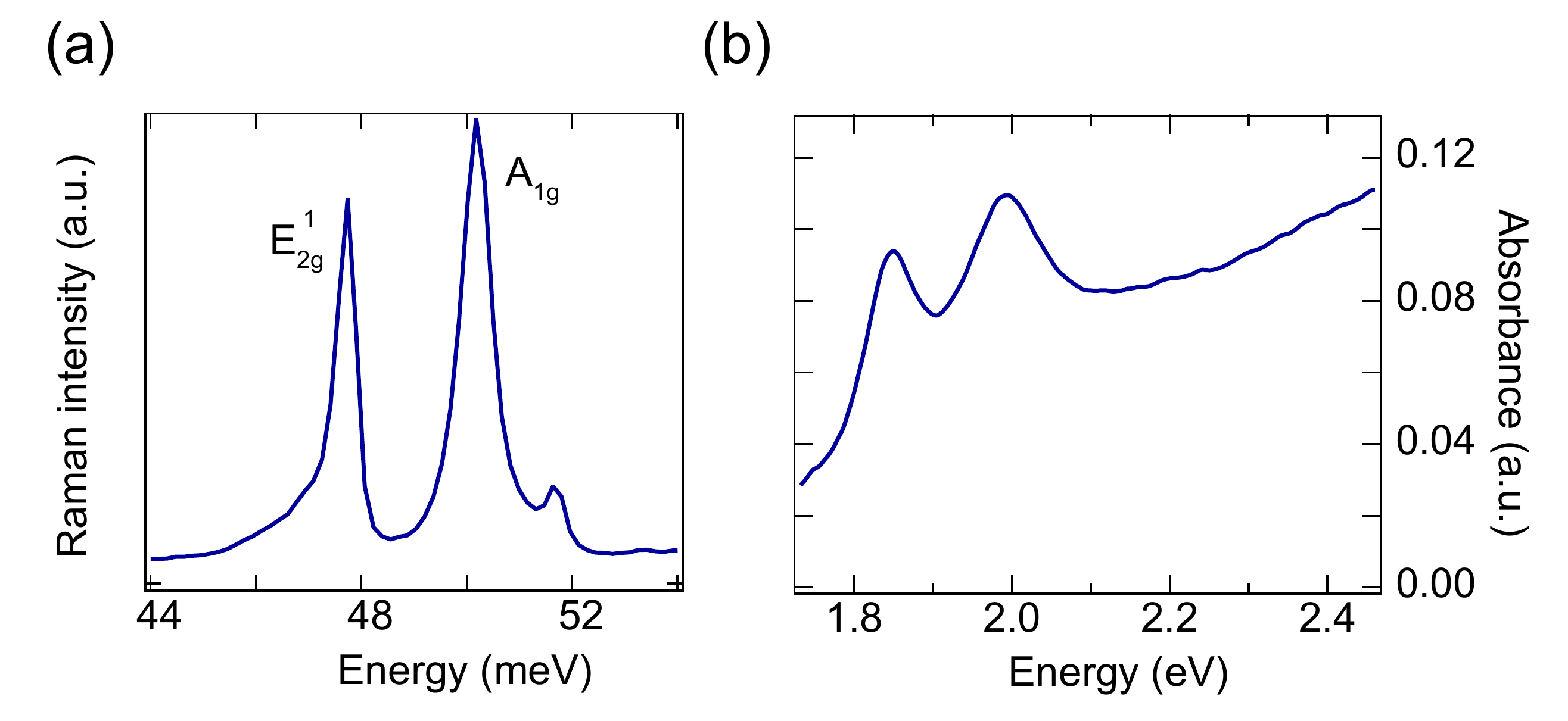}
		\caption{(a) Spontaneous Raman scattering spectrum of the sample. The difference between the energies of the $E_{2g}^1$ and $A_{1g}$ phonons indicates that the sample is a monolayer. (b) Room temperature steady-state absorption spectrum of monolayer MoS$_2$ showing the A and B exciton peaks.}
		\label{fig:FigS2}
	\end{center}
\end{figure}
\newpage

\begin{figure}[h]
	\begin{center}
		\includegraphics[width=0.7\columnwidth]{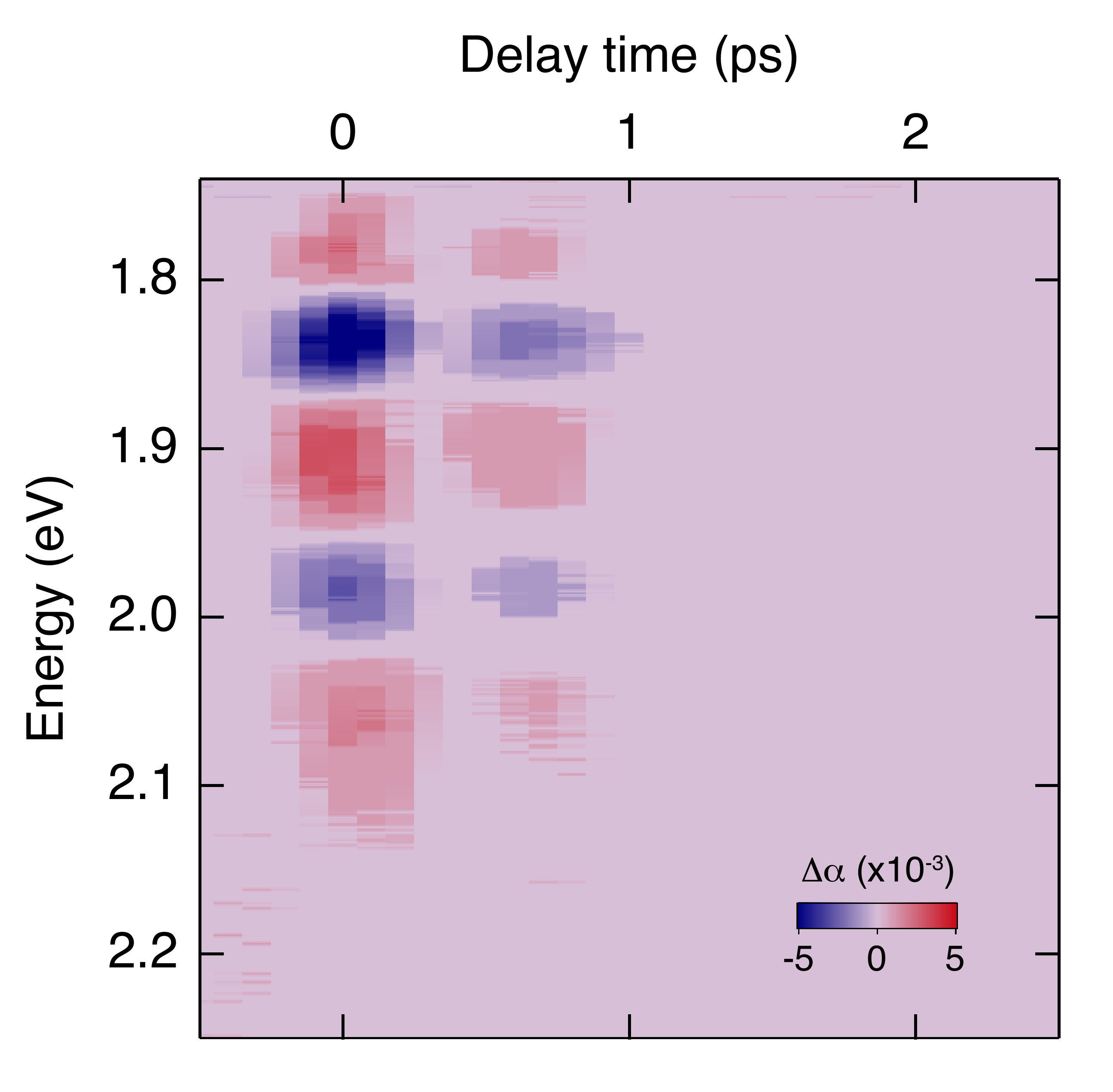}
		\caption{Color-coded map of the differential absorption ($\Delta\alpha$) as a function of the probe photon energy and the time delay between pump and probe. The map shows that no pump-probe signal is present after the THz excitation pulse is over.}
		\label{fig:FigS3}
	\end{center}
\end{figure}
\newpage

\begin{figure}
	\centering
	\includegraphics[width=0.6\columnwidth]{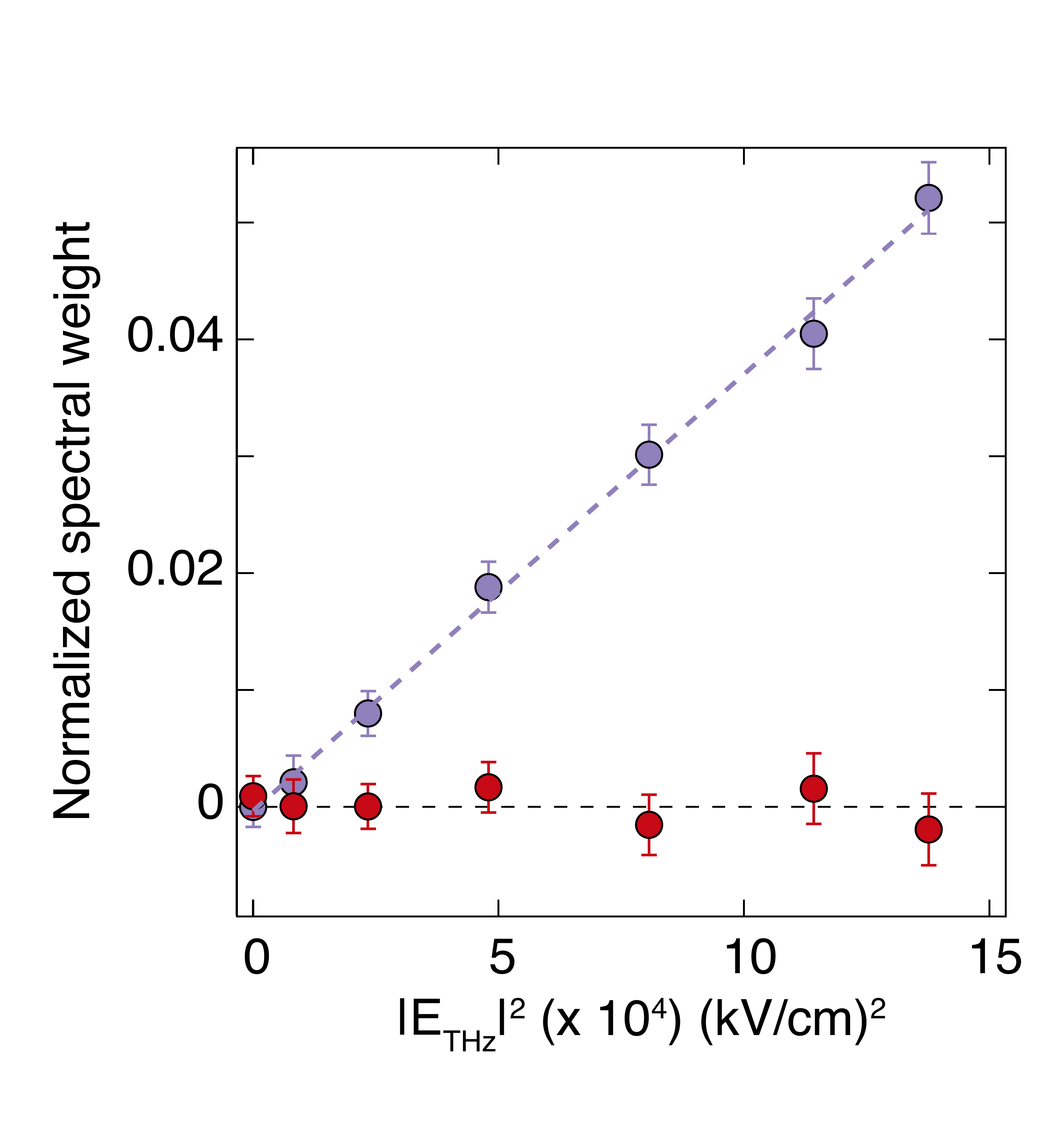}
	\caption{\label{FigS4}
		Quantitative analysis of the measured spectra in Fig. 3(a). The red data points are the values of the transient absorption areas normalized to zero-field absorbance area. The violet data points are the absolute values of the transient absorption areas (from 1.73-2.25 eV) normalized to the zero-field absorbance area. The vertical error bars are based on the experimental uncertainties given by the standard deviations from ten datasets taken under the same experimental conditions. 
	} 
\end{figure}
\newpage

\begin{figure}
	\centering
	\includegraphics[width=0.6\columnwidth]{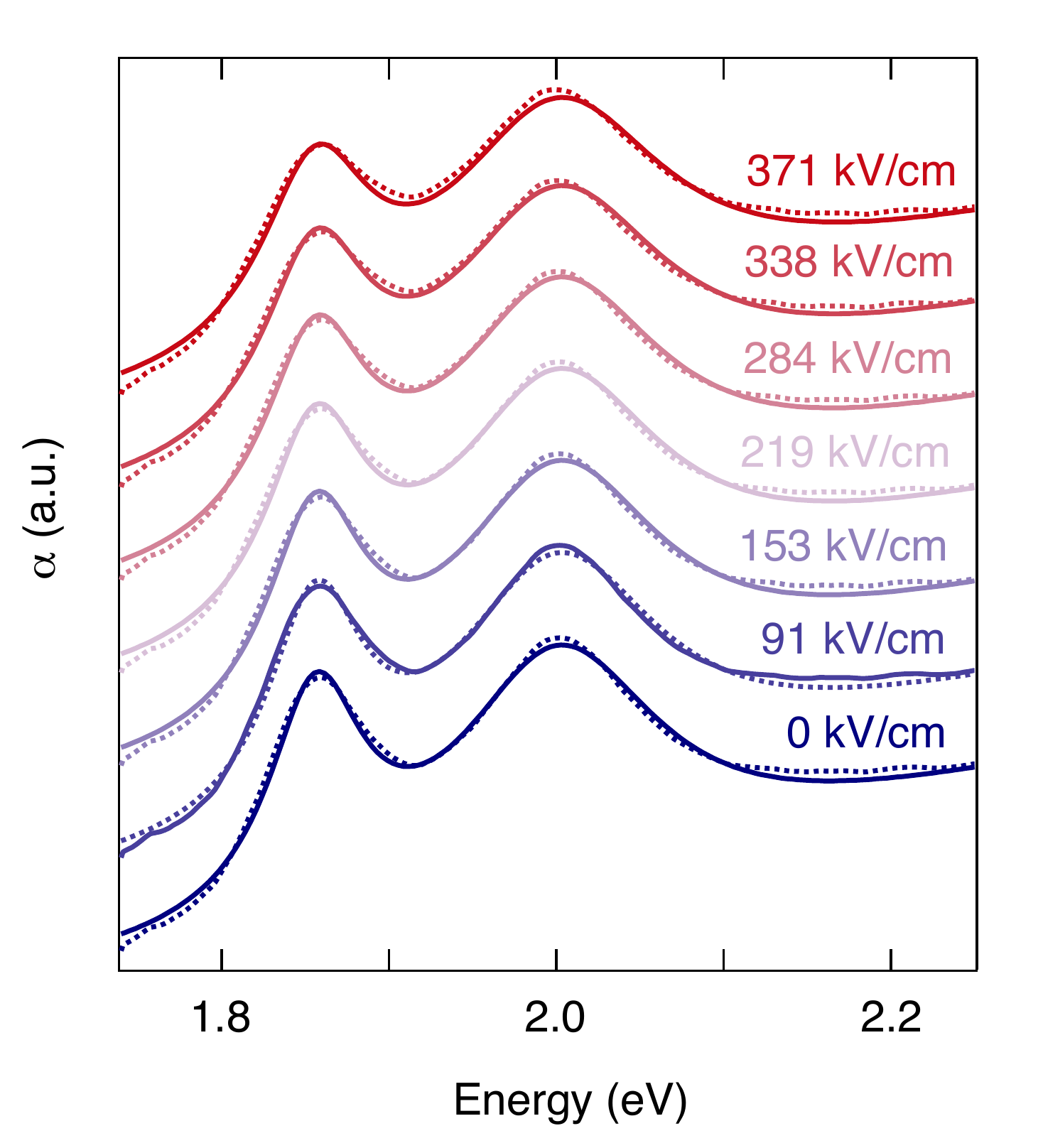}
	\caption{\label{FigS5}Comparison between the experimental absorption data (dotted lines) and the results of a phenomenological fit (solid lines) at different THz field strengths. The fit function comprises two Lorentz oscillators centered around excitons A and B, as well as a high-energy Lorentz oscillator whose parameters remain fixed at every THz field strength.
	} 
\end{figure}
\newpage

\begin{figure}
	\centering
	\includegraphics[width=0.6\columnwidth]{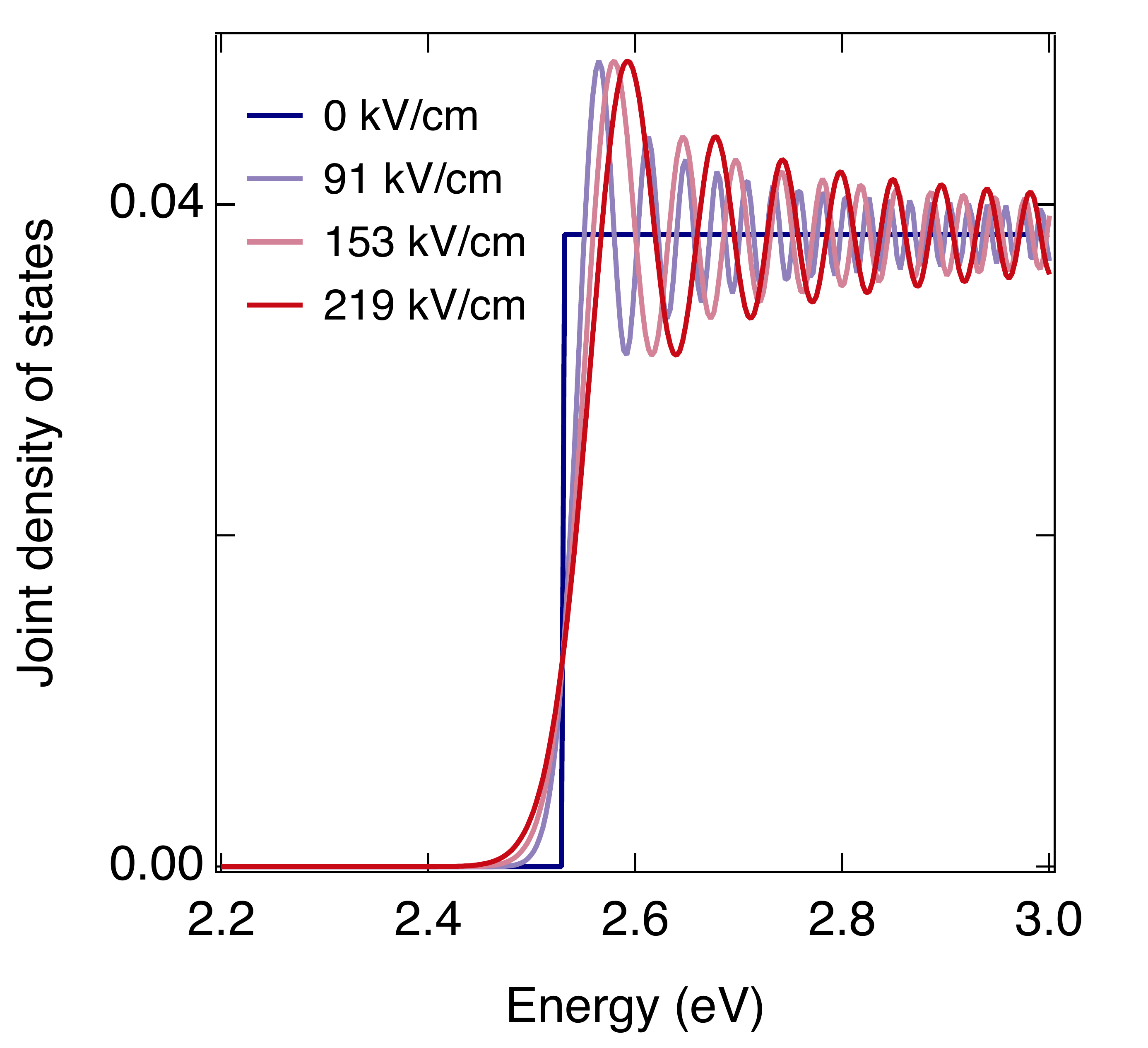}
	\caption{\label{FigS6}
		Joint density of states of a 2D crystal under the influence of a static electric field with each of the values indicated. The bands are assumed to have a parabolic energy-momentum dispersion relation.  The solid blue line shows the bandgap energy.
	} 
\end{figure}
\newpage

\begin{figure}
	\centering
	\includegraphics[width=0.6\columnwidth]{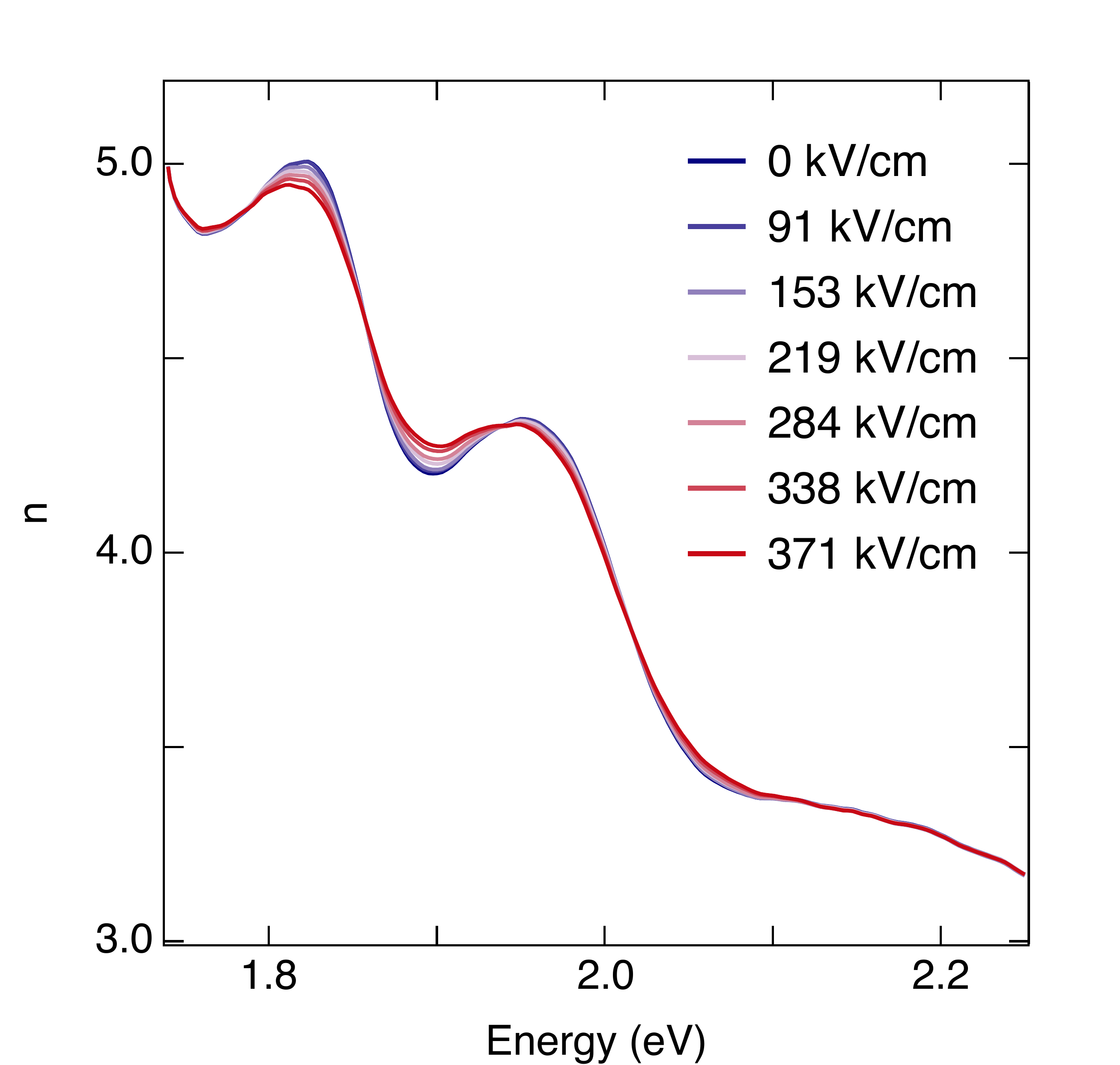}
	\caption{\label{FigS7}Real part of the refractive index of MoS$_2$ at zero pump-probe delay plotted in equilibrium (blue trace) and upon excitation with the THz field. The strength of the field is indicated in the label. The maximum variation obtained around 1.90 eV is $\sim$2.2$\%$ upon application of a THz field of 420 kV/cm.
	} 
\end{figure}
\newpage

\begin{figure}[h]
	\begin{center}
		\includegraphics[width=\columnwidth]{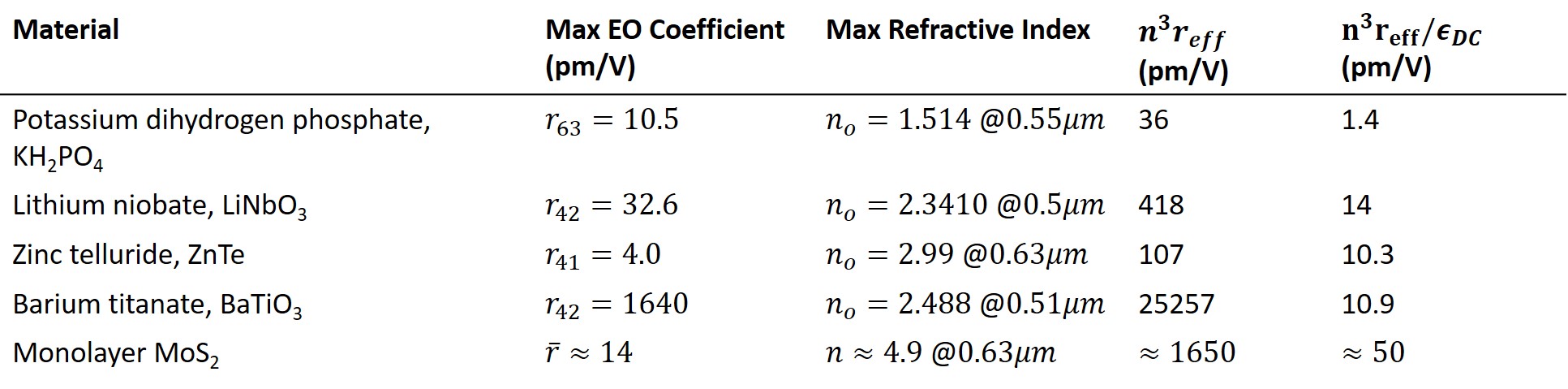}
		\caption{Comparison between the electro-optic properties of several materials and those of monolayer MoS$_2$.}
		\label{fig:FigS8}
	\end{center}
\end{figure}
\newpage
\clearpage

\providecommand{\noopsort}[1]{}\providecommand{\singleletter}[1]{#1}%
\providecommand{\latin}[1]{#1}
\providecommand*\mcitethebibliography{\thebibliography}
\csname @ifundefined\endcsname{endmcitethebibliography}
{\let\endmcitethebibliography\endthebibliography}{}

\end{document}